\documentclass[%
 amsmath,amssymb,
 aps,
12pt,
notitlepage
]{revtex4-1}

\usepackage{graphicx}
\usepackage{dcolumn}
\usepackage{bm}
\usepackage{hyperref}

\newcommand{\beq}{\begin{equation}}
\newcommand{\eeq}{\end{equation}}

\newcommand{\bea}{\begin{eqnarray}}
\newcommand{\eea}{\end{eqnarray}}

\newcommand {\ket}[1]{\vert{#1}\rangle}
\newcommand {\bra}[1]{\langle{#1}\vert}
\newcommand {\proj}[2]{\left\langle #1\vert #2\right \rangle}

\newcommand{\av}[3]{\left\langle #1 \left\lvert #2 \right\rvert #3 \right\rangle}

\newcommand{\e}[1]{e^{\left(#1\right)}}
\newcommand{\ih}{\frac{\imath}{\hbar}}

\begin{document}

\title{Entanglement dynamics in a system attached to self-interacting spinbaths }
\author{Nageswaran Rajendran}
 \email{nageswaran@chaos.physik.tu-dortmund.de}
\affiliation{%
Fachbereich Physik, Technische Universit\"at Dortmund, 44221 Dortmund, Germany. }%

\date{\today}
\begin{abstract}
Dynamics of quantum entanglement shared between system spins which are connected to thermal equilibrium baths is studied.  Central spin system comprises of the entangled spins, and is connected to baths and one of the bath has strong intra-environmental coupling. The dynamics between the system and baths are studied and inferred that that intra-envirnomental coupling guards against the system from the effects of the spin baths. 

\end{abstract}

\pacs{}

\maketitle


\section{\label{sec:level1}Introduction}
Wide ranges of open quantum dynamics are being studied in the implementations of quantum computing systems \cite{Deco-QtoC-2003-Zurek,Theory-OpenQSys-2002-Breuer}.  Decoherence is a natural quantum phenomena occurs when the system interacts with environments, depending on the coherence time, decoherence might lead to irreversible loss of information.  Basically, the environments can be independent harmonic oscillator baths \cite{QDiss-1983-Caldeira}, and  spinbath systems \cite{Theory-OpenQSys-2002-Breuer}.   Hence, we need to suppress the decoherence during the dynamics of the system, there are various methods, such as quantum error corrections \cite{QECC-GenNoise-2000-Knill}, dynamical decoupling \cite{UhrigSeq-2007-Uhrig}, decoherence free subspaces \cite{DFS_Lidar_1998} can be used in various implementation schemes.  However Tessieri and Wilkie suggested a model of central systems attached to a bath of spins interacting within themselves.   Tessieri-Wilkie model suggests that strong intracoupling between the spins in the baths can suppress the decoherence \cite{Tessieri-Wilkie-2003-Tessieri}, which is a necessity for the quantum computation and information implementations.  With the similar effect, entanglement sharing suppress the decoherence in a single spin self-interacting with the thermal(initially) baths \cite{EntSharing-Deco-2005-Dawson} and strong decoupling effects on the disentanglement in the system against the variations in coupling constants between the baths \cite{Tessieri-Wilkie-2007-Olsen}.  The model consists of a central system of spins, which is connected with $N$ number of spin baths with an intra-coupling strength $\lambda$.  Description of this type of open quantum system can be represented by the Hamiltonian.

\section{\label{sec:twmodel}Model}

A pair of spin-1/2 system is taken as in an entangled state $\ket{\Psi_0}$, initially, and it is connected to the two system of baths, in which first bath has a coupled spin pair and another bath has one spin.  Dynamics of the system is studied under the conditions of ferromagnetic and antiferromagnetic interaction shared among the spins.

\bea
H=H_S+H_B+H_{SB}
\label{eq:Ham}
\eea

Where, $H_S$, $H_B$ and $H_{SB}$ are denoted as the Hamiltonian system, bath and system-bath description, respectively.   The Hamiltonian of the $k^{th}$ spin in the central system is given as

\bea
H_{S}^k &= & \frac{\hbar \omega_{S}}{2} \sigma_{z}^{k}+ \beta \sigma_{x}^{k} 
\label{eq:HamSys}
\eea
where $\omega_k$ is the frequency of the system and $\sigma_{x,y,z}$ are Pauli operators; $\beta$ represent reducibility of the Hamiltonian, in other words, spin ups and downs which form a even and odd subspaces formed by the basis states of the system,

\bea
\beta 
\begin{cases}
\neq 0, & \text{Hamiltonian becomes irreducible;}\\
= 0,		&\text{reducible Hamiltonian.}
\end{cases}
\eea
Using the natural units, $\hbar$, $k_{B}$ are set to unity.  Each of the bath description is represented as

\bea
H_B^l=\sum_i \frac{\omega_i}{2}\sigma_z^i+ \beta\sum_i\sigma_{x}^i+\lambda\sum_{\substack{ i,j\\i\neq j}} \sigma_x^i\sigma_x^j
\label{eq:HamB}
\eea

\bea
\lambda =
\begin{cases}
[0, n], & n \in \mathbb{R}, \text{for antiferromagnetic case}\\
[-n, 0], & \text{ferromagnetic case.}
\end{cases}
\eea
$\lambda$ is the spin-spin coupling in each of its bath.  And the system-bath description is 
represented as

\bea
H_{SB}= \lambda_0 \sum_l \sigma_x^{S_k} \sigma_x^{B_l}
\label{eq:HamSB}
\eea

The initial state of the open system can be described by $\rho(0)=\rho_S(0)\otimes\rho_B(0)$,
the initial state of the bath is in thermal equilibrium and it is in the form, 

\bea
\rho_B(0)=\frac{exp(-H_B/kT)}{Tr(exp(-H_B/kT))}
\label{eq:rhoB}
\eea 

 Evolution of the entire system under its Hamiltonian is represented  by the interaction picture.

\bea
\rho(t)=e^{-iHt} \rho(0) e^{iHt}
\label{eq:evol}
\eea  
  \section{Entanglement evolution}
  
As an entanglement measure concurrence has been used for this case, which built over the entanglement of formation, and may be the generalized measure for the class of arbitrary mixed entangled states \cite{Concurrence-1998-Wootters}.  Concurrence of an entangled state $\rho$ is given by 

\begin{equation}
C[\rho]=\max\{0, \lambda_1-\sum_{i=2}^3 \lambda_i\}
\label{eq:C}
\end{equation}

As, $\lambda_i$ is the eigenvalue of the matrix $\rho \tilde\rho$, and $\tilde\rho$ is defined as ($\sigma_y\otimes\sigma_y) \rho (\sigma_y\otimes\sigma_y)$, and $\lambda_i$ are arranged in decreasing order.

  The system has two spins, which are taken as an arbitrary entangled pair $\ket \Psi=\exp{(-i \delta/2)} \cos{(\alpha/2)} \ket{00} + \exp{(i \delta/2)} \sin{(\alpha/2)} \ket{11}$, as our state.   With the different set of cases are calculated for the initial states
  
    \bea
\ket{\Psi(0)}=
\begin{cases}
 \frac{1}{\sqrt2} \ket{00} +  \frac{1}{\sqrt2} \ket{11} &\\
 \frac{\sqrt{3}}{2}e^{-i \pi/8} \ket{00} + \frac{1}{2}e^{i \pi/8} \ket{11}&\\
\frac{\sqrt{3}}{2}e^{-i \pi/8} \ket{00} + \frac{1}{2}e^{i \pi/8} \ket{11}+\epsilon \ket{01}.
\end{cases}
\label{eq:SysInit}
\eea

     The system is connected to all the bath through the coupling with the strength of $\lambda_0$.  The dynamics of the system coupled to spin baths, which are also inter connected through the (anti-)ferromagnetic coupling.  The system is connected to two baths, which are independent, and first bath has two spins interacting within themselves  with a coupling strength $\lambda$ which varies from $\lambda\in[-2,2]$ and in some cases stronger coupling strength is considered for comparative purposes.  And the second bath has a single spin.

From the equations \ref{eq:HamSys}, \ref{eq:HamB} and \ref{eq:HamSB}, the system Hamiltonian is defined as,
\bea
 H_S&= &\left( \frac{ \omega_{s}}{2} \sigma_{z}^{(1)}+ \beta \sigma_{x}^{(1)} \right)\otimes \left(\frac{\omega_{s}}{2} \sigma_{z}^{(2)}+ \beta \sigma_{x}^{(2)}\right).
 \label{eq:Hs}
\eea

Hamiltonian of the first spin bath with the intra-bath coupling $\lambda$ is given by

\bea
H_B^{(1)} &=& \frac{\omega_{b1}^{(1)}}{2}\sigma_z^{(1)}+\frac{\omega_{b1}^{(2)}}{2}\sigma_z^{(2)}+ \beta\sigma_{x}^{(1)}+ \beta\sigma_{x}^{(2)}+\lambda \sigma_x^{(1)}\sigma_x^{(2)}, 
\label{eq:Hb1}
\eea
and also the Hamiltonian of the second bath with a single spin is given by
\bea
H_B^{(2)} &=&\frac{\omega_{b2}^{(1)}}{2}\sigma_z^{(1_{b2})}+ \beta\sigma_{x}^{(1_{b2})}.
\label{eq:Hb2}
\eea

The system and bath Hamiltonian with the coupling strength $\lambda_0$ is written as
\bea
H_{SB}&=& \lambda_0( \sigma_x^{s}\sigma_x^{B_1}+\sigma_x^{s}\sigma_x^{B_2})
\label{eq:Hsb}
\eea

For the sake of brevity, let us assign the values $\omega_s=0.7$, as the system has lesser energy as the system is in low temperature.   And $\omega_B^i=1$ and $\beta=0.01$ and considering the system lies in the low temperature $k_B T=0.02$

The density matrices of the system and first and second bath can be represented as 

\bea
\rho_0 &=& \rho_s^{(1,2)} \otimes \rho_{B_1}^{(3,4)} \otimes \rho_{B_2}^{(5)} \\
\rho(t) &=& \exp(-iH t) \rho_0 \exp(iHt)
\eea

$\rho(t)$ is traced over the baths, which gives $Tr_{3,4,5}(\rho(t))=\rho_s^{1,2}$, the reduced density matrices of the system.

  The quantity of entanglement shared between the system spins is affected through system-bath interaction and  intra-bath coupling among the spins in the bath.  Therefore, the concurrence of the evolved state may be viewed as a function of coupling parameter among the spins in the bath and the evolution of the state with respect to time.   Hence the concurrence of the state of the system $C(\rho_s^{1,2})$ is calculated using the \ref{eq:C}  numerically, for the different initial states as mentioned  in \ref{eq:SysInit}.

 \subsection{Entangled state}
 
\begin{figure}[hbtp]
   \hfill
   \begin{minipage}[t]{.45\textwidth}
(a) \includegraphics[scale=0.4]{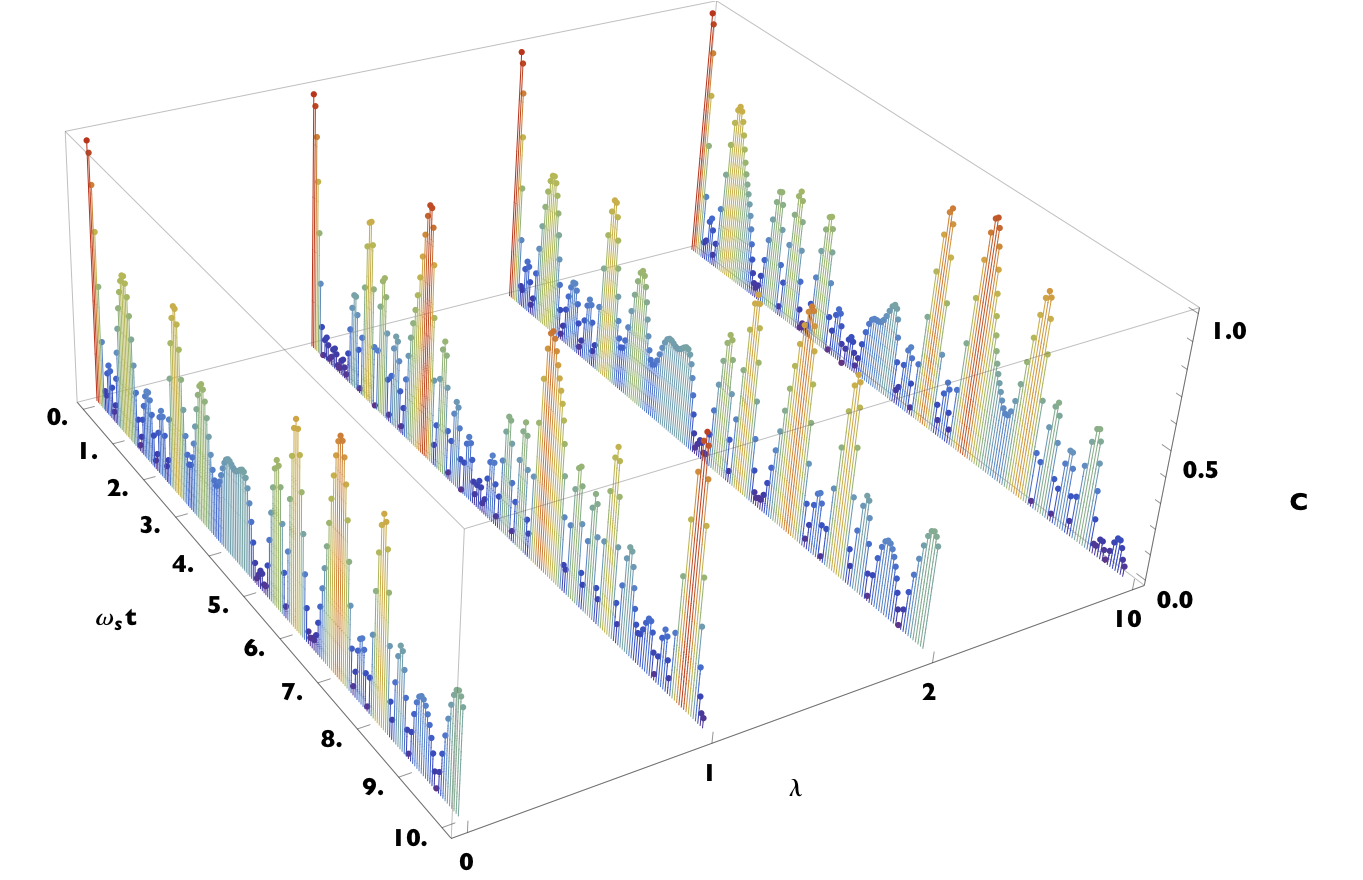}
   \end{minipage}
 \hfill
   \begin{minipage}[t]{.45\textwidth}
(b) \includegraphics[scale=0.4]{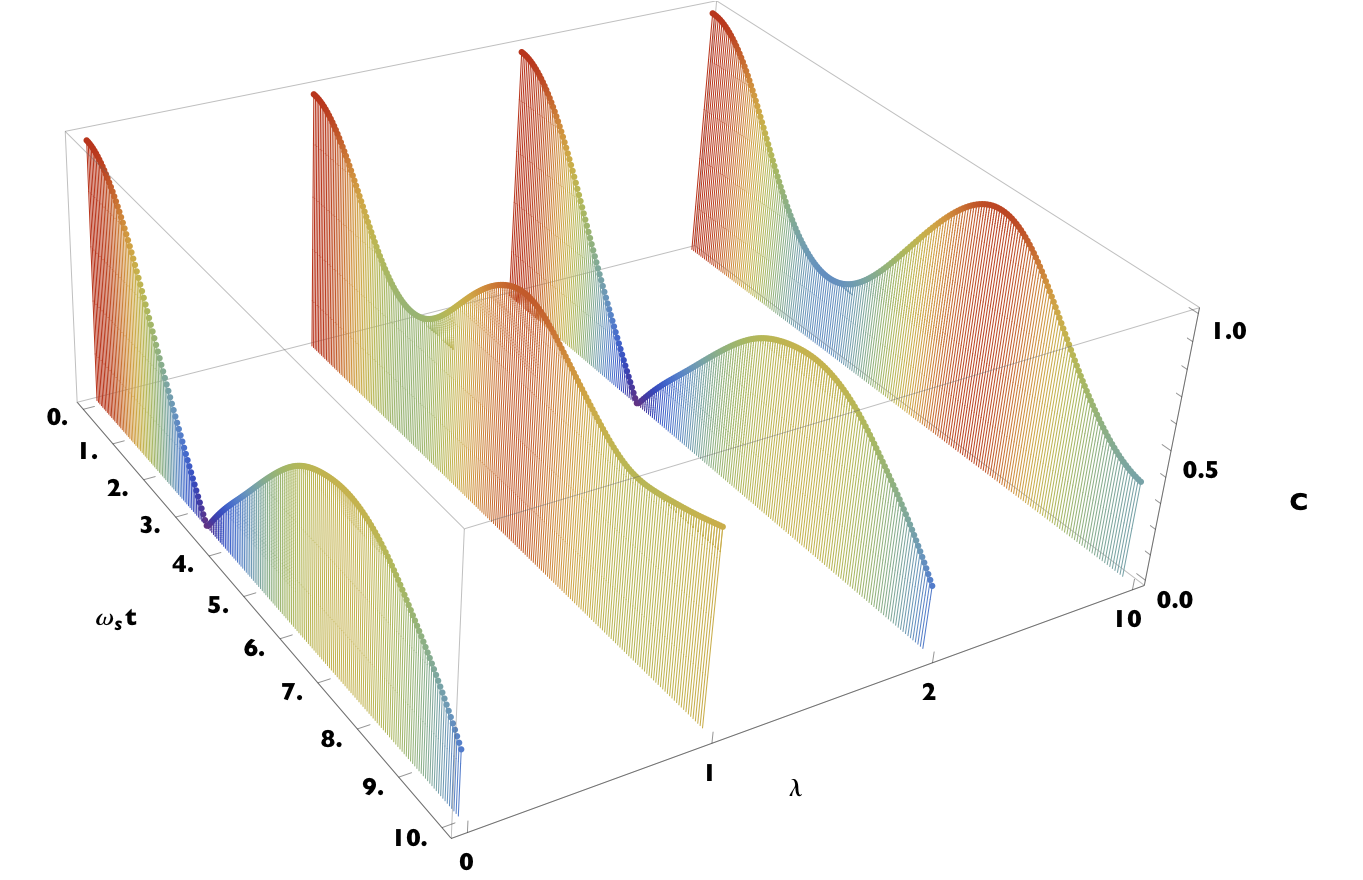}
\label{ct-bell}
   \end{minipage}
  \hfill
   \caption{Concurrence plotted for the state $\ket\Psi= \frac{1}{\sqrt2} (\ket{00} + \ket{11} )$ various coupling strength $\lambda=0,1,2,10$ and the system bath coupling $\lambda_0=0.1$ fig.(a) and $\lambda_0=1$ fig(b)}  \label{partent}
 \end{figure}

 A maximally entangled state is taken as the initial state for the first case.  Hence the concurrence for the entangled state must read from unity.     Both at Antiferromagnetic and ferromagenetic regions, the system tend to behave similar manner, while comparing the concurrence of the system.   When the intra-bath coupling goes to zero, the system,
 When the intra-environmental coupling gets stronger, effects on the eigenstates of the systems lesser compared to the weaker coupling cases.   From the fig\ref{ct-bell}(a) and (b), we could see for the higher intra-bath coupling the, the system tries to revoke its entanglement as the system and bath evolves.

System-bath coupling has additional effects when that fluctuate between 0.1 and 1.  For the stronger $\lambda_0=1$ system-bath coupling, the eigenstates of the spinbaths gets coupled to the system, which decreases the entanglement of the state of the system.  Nevertheless, the evolution of the system tells that the decoherence may be effectively handled.

 \subsection{Partially entangled state}
 In the case of partially entangled state as the input state, $\ket\Psi= \frac{\sqrt{3}}{2}e^{-i \pi/8} \ket{00} + \frac{1}{2}e^{i \pi/8} \ket{11}$, the concurrence of the initial state is $0.6$, and the entanglement dynamics follows the same pattern as it happened in maximally entangled state during the evolution.
 
 Entanglement measure for the system at the initial state is calculated as 0.6.
   \begin{figure}[hbtp]
 \hfill
   \begin{minipage}[t]{.5\textwidth}
(a) \includegraphics[scale=0.5]{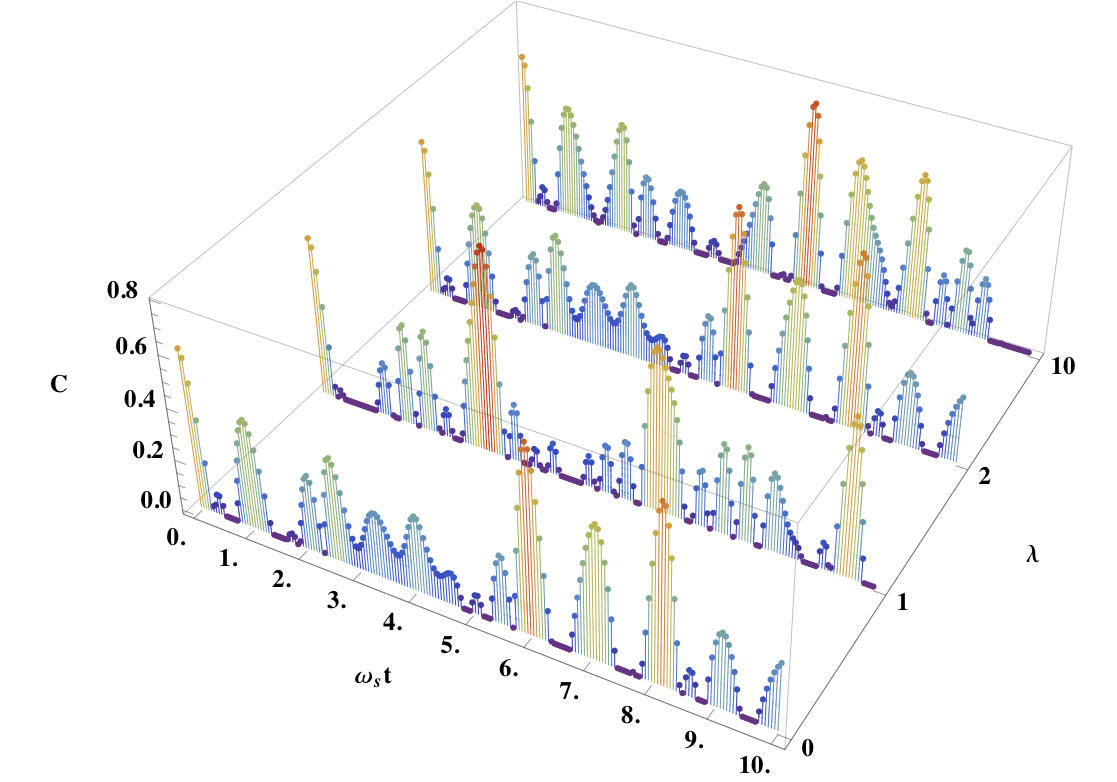}
   \end{minipage}
  \hfill
   \begin{minipage}[t]{.40\textwidth}
(b) \includegraphics[scale=0.4]{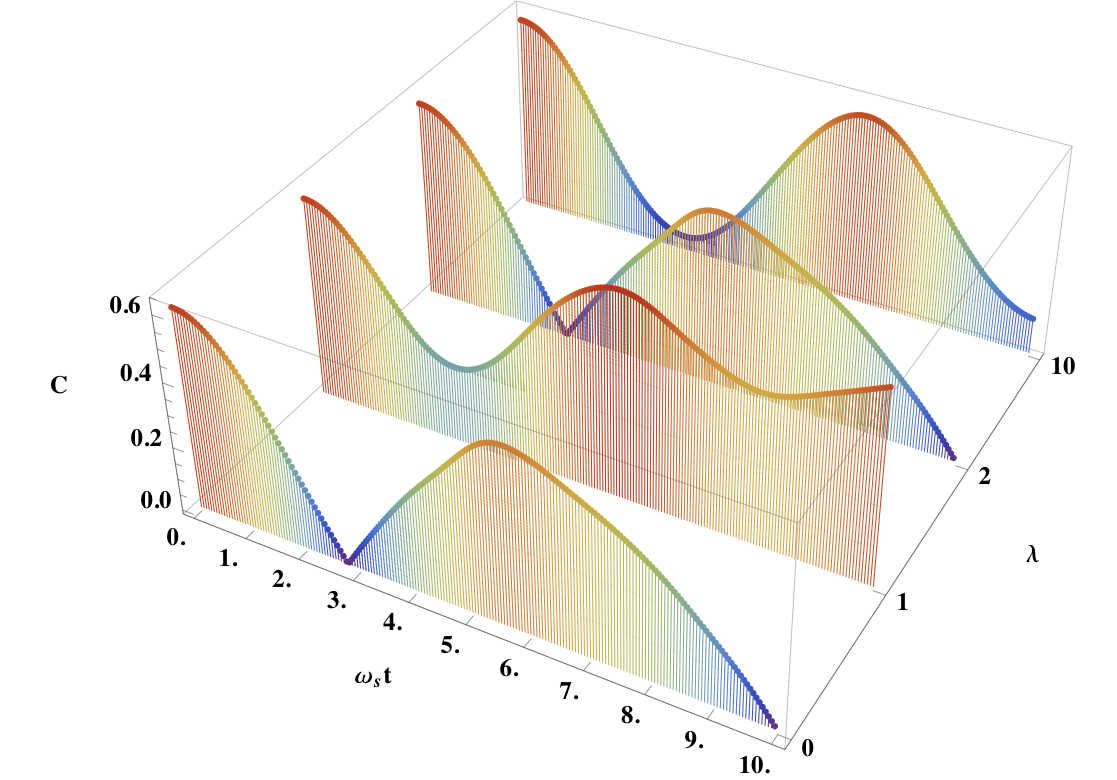}
   \end{minipage}
  \hfill
  \label{ct-partent}
  \caption{Concurrence plotted for the state $\ket\Psi=\frac{\sqrt{3}}{2}e^{-i \pi/8} \ket{00} + \frac{1}{2}e^{i \pi/8} \ket{11}$ various coupling strength $\lambda=0,1,2,10$ and the system bath coupling $\lambda_0=0.1$ and $\lambda_0=1$ }\label{partent}
 \end{figure}

%
%
 
The partially entangled state with additional elements added to the entangled state $\ket\Psi= \frac{\sqrt{3}}{2}e^{-i \pi/8} \ket{00} + \frac{1}{2}e^{i \pi/8} \ket{11}+\epsilon (\ket{01}+\ket{10}).$
 
\begin{figure}[hbtp]
   \hfill
   \begin{minipage}[t]{.45\textwidth}
(a) \includegraphics[scale=0.4]{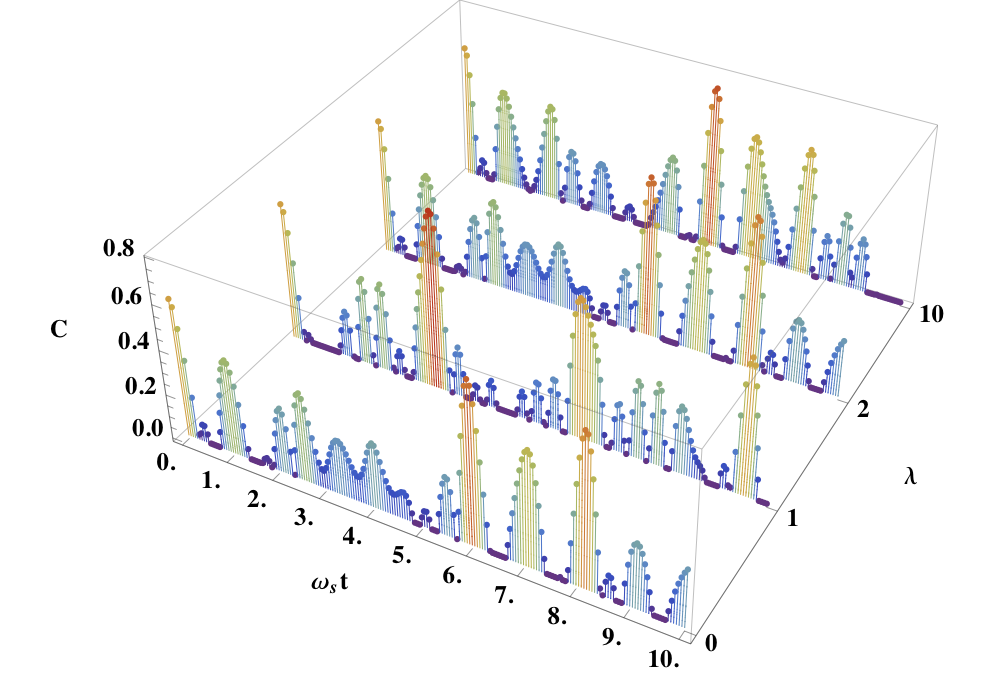}
   \end{minipage}
  \hfill
   \begin{minipage}[t]{.45\textwidth}
(b) \includegraphics[scale=0.4]{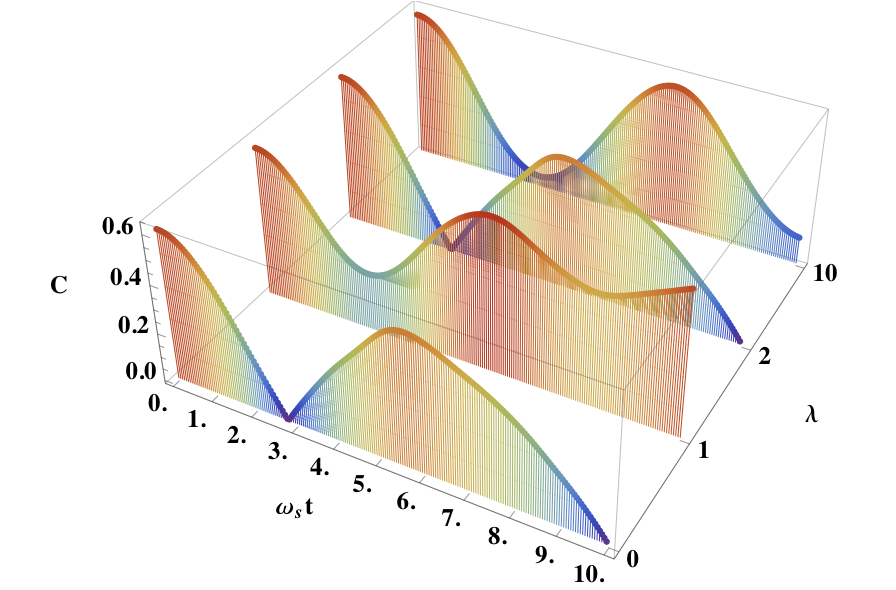}
   \end{minipage}
   
   \caption{Concurrence plotted for the state $\ket\Psi=\frac{\sqrt{3}}{2}e^{-i \pi/8} \ket{00} + \frac{1}{2}e^{i \pi/8} \ket{11}+\epsilon \ket{01}$ various coupling strength $\lambda=0,1,2,10$ and the system bath coupling $\lambda_0=0.1$ and $\lambda_0=1$ }\label{partent-error}
   \end{figure}
   
   Similar to the maximally entangled cases, the system retrieves its concurrence during the evolution; even for the partially entangled and entangled mixed states, the system seems to be robust against its interaction with two of its baths.
 
However, effect of the system-bath coupling, for $\lambda_0=1$ seems to affect the system heavily.   To see the effects of quantum Zeno dynamics during the evolution for the higher system-bath coupling could be interesting.  
%

 \section{Frequent measurements}
 Quantum Zeno effects can be viewed quantum systems by doing frequent measurements   \cite{QZE-1977-Sudarshan}.  Applying frequent measurements on the system decouples the system from its environment.    If the initial state is in $\rho^{(1,2)}=\ket{\psi^{(1)}}\bra{\psi^{(2)}}$, which undergoes the evolution $U(t_k)=\exp{i H t_k}$, where $t_K$ is the survival time, $t > t_k$, the state will start to decay.   Hence frequent  projective measurement within the interval $t_k$ will lead the state stay in the state.  Therefore, the survival probability is given by,
 
 \begin{eqnarray}
P_{zeno}=\|\proj{\psi_0}{\psi(t_N)}\|^2=\left(\prod_{k=1}^N \av{\psi_0}{\e{\ih t_k(H)}}{\psi_0}\right)^2
\end{eqnarray}

Where $N$ is the iterated measurements.  When the selective projective measurement is made the probability is proportional to $ \cos^{(2N)}(\delta/2)$.  In our case, $|Tr(e^(-iHt_k) \rho_s)|^2$ calculated to compute the Zeno probability.

When the system has lesser coupling strength with the baths,  the entanglement of the system slowly decreases during the free evolution.   To suppress the decay of the state of the system, number of frequent measurements are applied during the evolution $t_k=2 \pi /\lambda_0$ during the free evolution.

\begin{figure}[hbtp]
   \hfill
  (a) \begin{minipage}[t]{.45\textwidth}
\includegraphics[scale=0.45]{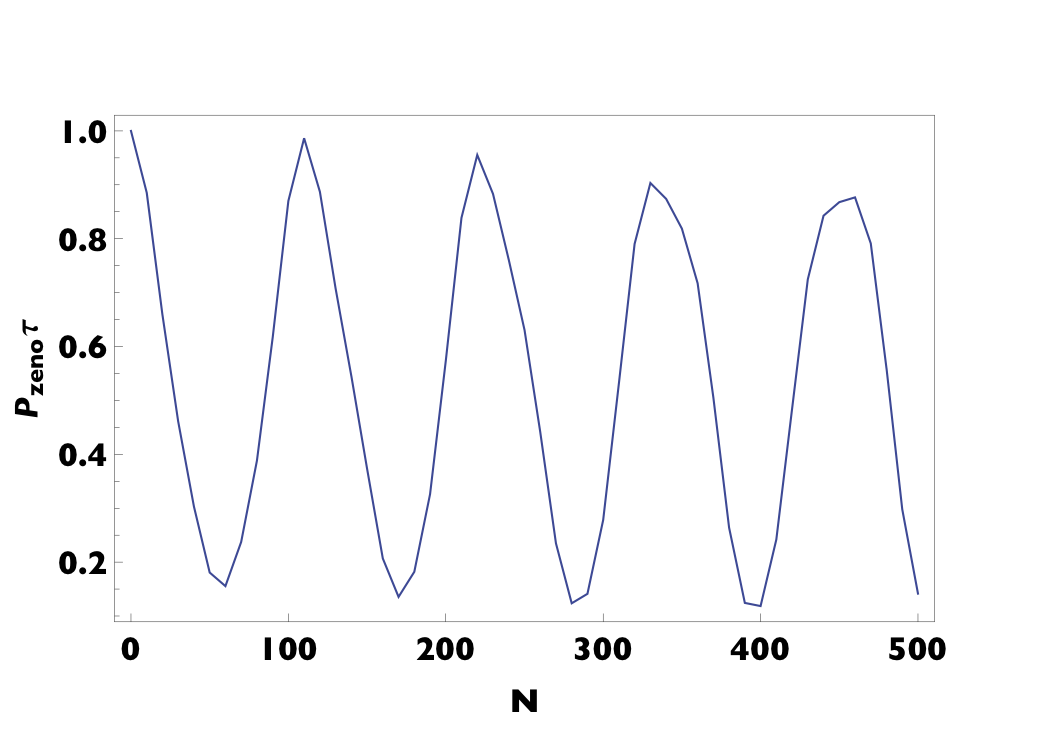}
   \end{minipage}
 \hfill
  (b) \begin{minipage}[t]{.45\textwidth}
\includegraphics[scale=0.45]{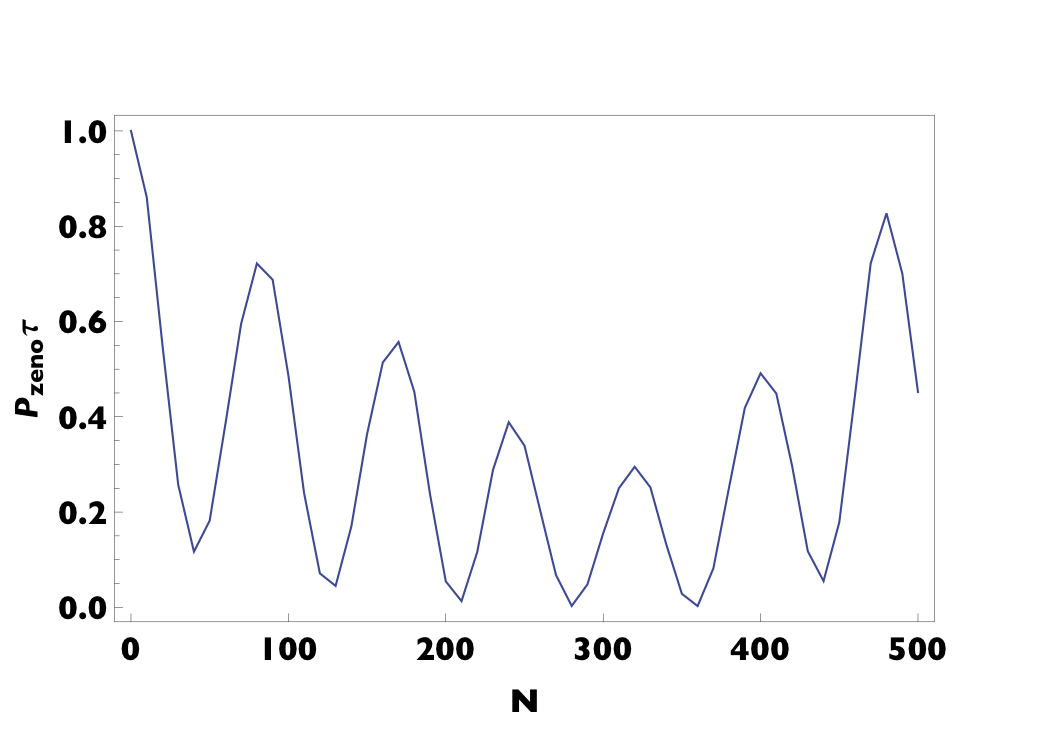}
   \end{minipage}
  \hfill
   \caption{Survival probability of the state during the evolution for the $\lambda_0=0.1,1$}
   \label{zenoprob}
 \end{figure}  


With the Zeno measurements, we can see \ref{zenoprob} the survival probability of the state increases as the number of measurements increases even in the strong system-bath coupling.

\section{Conclusion}
By having the spinbaths with strong intra-bath coupling strength,  robustness of the state of the system may be preserved.   The entanglement measure of the system goes down when the system gets interacted to the baths, which are in thermal equilibrium initially, and the decreasing effect may be due to `inertia' of the baths, as we could also see when intra-bath coupling nears to zero, the rate of change of concurrence is relatively faster.    The system retrieves its initial amount of entanglement, as the system progresses to evolve under its Hamiltonian. 

As we have seen the entanglement goes down, when the system undergoes relatively smaller interaction, i.e, weakly coupled with the baths.  The system has been set to undergo into the frequent measurements to view the robustness of the system.   During the frequent measurements with various number of iterations. 

The interaction between the system and bath is playing an important role in the system-bath dynamics, apart from the strong coupling between the intra-bath spins.   This could  be an interesting inference for the experimental implementations.   

\bibliography{whol}
\end{document}